\documentclass[twocolumn,aps,tightenlines,floats,floatfix]{revtex4}

\def\be{\begin{equation}}
\def\ee{\end{equation}}
\def\ba{\begin{eqnarray}}
\def\ea{\end{eqnarray}}

\usepackage{graphics}
\usepackage{graphicx}
\usepackage{epsf} 
\usepackage{amsmath}
\usepackage{amssymb}
\usepackage{bbold}
\def\sfrac#1#2{{\textstyle \frac{#1}{#2}}}

\begin{document}

\phantom{0}
\hspace{5.5in}
\parbox{1.5in}{ \leftline{JLAB-THY-07-700}}

\vspace{-1in}

\title
{\bf Fixed-axis polarization states: covariance and comparisons}

 \author{Franz Gross$^{1,2}$, G. Ramalho$^{2,3}$ and M.T. Pe\~na$^{3,4}$
\vspace{-0.1in}  }

\affiliation{
$^1$College of William and Mary, Williamsburg, Virginia 23185 \vspace{-0.15in}}
\affiliation{
$^2$Thomas Jefferson National Accelerator Facility, Newport News, 
VA 23606 \vspace{-0.15in}}
\affiliation{
$^3$Centro de F{\'\i}sica Te\'orica e de Part{\'\i}culas,
Av.\ Rovisco Pais, 1049-001 Lisboa, Portugal
\vspace{-0.15in}}
\affiliation{
$^4$Department of Physics, Instituto Superior T\'ecnico, 
Av.\ Rovisco Pais, 1049-001 Lisboa, Portugal}

 \begin{abstract}
 
 Addressing the recent criticisms of  Kvinikhidze and Miller, we prove that the spectator wave functions and currents based on  ``fixed-axis'' polarization states (previously introduced by us) are  Lorentz covariant, and find an explicit connection between them and conventional direction-dependent polarization states.  The discussion shows explicitly how it is possible to construct pure $S$-wave models of the nucleon.
 \end{abstract} 
 \phantom{0}
\vspace*{0.85in}  

 \maketitle
\section{Background}

\vspace{-0.15in}

In Ref.\ \cite{Gross:2006fg} (referred to as GRP) we propose a new  definition of vector polarization states, which we refer to as ``fixed-axis'' states.   Using fixed-axis states it is comparatively easy to construct phenomenological wave functions with a pure $S$-wave structure and use them to investigate the role of angular momentum components in determining the structure of the nucleon form factors.  A detailed discussion of this phenomenology is presented in GRP.

Kvinikhidze and Miller \cite{Kvinikhidze:2007qq} (referred to as  KM) claimed that the definition of the vector  polarization states used in an earlier version of GRP (available from the eprint archives \cite{Gross:2006fgv1}) were not covariant.  The discussion presented in \cite{Gross:2006fgv1} was very sketchy, and therefore understandbly open to misinterpretation.   The purposes of this paper are (i)  to present this new physics in a systematic way, and (ii) to answer the  objections of KM. 

The remainder of this paper is divided into three sections.  In Sec.~II, 
building on the physical intuition introduced in GRP, we present a coherent 
discussion of wave functions and corresponding electromagnetic currents 
based on the new {\it fixed-axis\/} polarization vectors and show that they 
are covariant.  This 
work is previously unpublished, but was available to KM as a private 
communication.  In Sec.~III we compare wave functions and electromagnetic 
currents defined with fixed-axis states to wave functions and electromagnetic currents defined with 
helicity states.   We show, for the first time, how the wave functions and electromagnetic currents defined using the two
different bases are related, and what this means.
Finally, Sec.~IV draws some conclusions.

\section{Covariance of fixed-axis matrix elements}

\subsection{Fixed-axis polarization states for a bound axial vector diquark}

In GRP  a basis of fixed-axis  polarization vectors  are introduced.  These are intended to be used only for vector particles that are constituents of a bound state of four-momentum $P$.    In analogy with the construction of helicity state polarizations, fixed-axis polarizations are constructed through a sequence of steps.  The polarization of the diquark is first defined in the rest frame of the nucleon by  expanding the vector in terms of four-vectors $\varepsilon_{0}(\lambda)$ with angular momentum projections $\lambda=\{1,0,-1\}$ along the $z$ axis
\be
\varepsilon^\mu_{0}(\pm)=\mp\sfrac{1}{\sqrt{2}}\left[\begin{array}{r} 0 \cr 1 \cr \pm i \cr 0\end{array}\right]\, , \qquad 
\varepsilon^\mu_{0}(0)=\left[\begin{array}{r} 0 \cr 0 \cr 0 \cr 1\end{array}\right]\, . \label{eq.1}
\ee
This provides a polarization basis for the diquark that is independent of the magnitude or direction of the momentum of the diquark.  This definition has a number of advantages for the phenomenological treatment of quark models, and these will be discussed elsewhere.

When the bound state is at rest the fixed-axis can be chosen to be in any direction.  When the bound state is in motion, we choose the fixed-axis to be in the direction of
the  three-momentum of the bound state.  Choosing this to be the $z$ axis, the bound state four-momentum is $P=\{{\cal E}_p,0,0,p\}$, where ${\cal E}_p=\sqrt{M^2+p^2}$ and $M$ is the mass of the bound state.  The polarization vectors in this frame are obtained by boosting (\ref{eq.1}), and become
\ba
\varepsilon^\mu_p(\pm)&=&Z^\mu{}_\nu\, \varepsilon^\nu_{0}(\pm)=\mp\sfrac{1}{\sqrt{2}}\left[\begin{array}{r} 0 \cr 1 \cr \pm i \cr 0\end{array}\right]\, , \nonumber\\ 
\varepsilon^\mu_p(0)&=&Z^\mu{}_\nu\, \varepsilon^\nu_{0}(0)=\frac{1}{M}\left[\begin{array}{c} p \cr 0 \cr 0 \cr {\cal E}_p\end{array}\right]\, , \label{eq.4}
\ea
where the subscript on the fixed-axis polarization vectors denotes the momentum of the bound state.  Note that, for boosts in the $z$ direction,  there are no Wigner rotations for fixed-axis polarization vectors.

These polarization vectors satisfy the orthonormality conditions
\be
\varepsilon_p^*(\lambda')\cdot\varepsilon_p(\lambda)=-\delta_{\lambda\lambda'}\, , \label{eq.9b}
\ee
but  differ from helicity states in one important respect.  Helicity-state polarizations are orthogonal to $k$ (the four-momentum of the diquark) while fixed-axis polarizations are orthogonal to $P$
\ba
P\cdot \varepsilon_p(\lambda)=0\, . \label{eq.10b}
\ea
The fixed-axis polarizations are unconventional in the sense that they describe the polarization of the diquark without reference to the direction of the three-momentum of the diquark ${\bf k}$.   This is a common convention in nonrelativistic physics, but not used before (as far as we know) in relativistic physics.

\subsection{The spectator wave function}

In the spectator theory \cite{Gross69}, the structure of a two-body bound state enters through a matrix element of a field operator  between one constituent (which is on-shell) and the bound state (also on-shell).  This idea was first introduced by Blankenbecler and Cook  in 1960 \cite{BC}, and has been used extensively to describe the relativistic  structure of the deuteron \cite{Buck:1979ff}.   In that application the relativistic deuteron wave function is related to the covariant $d\to np$ vertex function $\Gamma$ by the relations
\ba
\Psi_{\lambda_1\beta;\lambda}(p,P)&=&\left<p_1,\lambda_1|\psi_\beta(0)|P,\lambda\right>\nonumber\\
&=&\xi_\mu(\lambda)\big[S_F(p_2) \Gamma^\mu(p_1,P){\cal C}\big]_{\beta\beta'}\bar{u}_{\beta'}^{\rm T}({p}_1,\lambda_1)\nonumber\\
&=&\sum_{\lambda_2,\rho}\psi^\rho_{\lambda_1\lambda_2,\lambda}\,u^\rho_\beta({p}_2,\lambda_2)\, . \label{eq5a}
\ea
where $p=\sfrac12(p_1-p_2)$, $P=p_1+p_2$ ($p_1$ and $p_2$ are the four-momenta of the two nucleons with $p_1^2=M^2$), $S_F$ is the propagator of the off-shell nucleon, and $\xi(\lambda)$ is the polarization of the deuteron with helicity $\lambda$, and the sum is over the polarizations and $\rho$-spins (where $\rho=+$ are positive energy $u$ spinors and $\rho=-$ are negative energy $v$ spinors)  of the off-shell nucleon.  For more discussion, see Ref.~\cite{Buck:1979ff}.

Here we use similar ideas to describe the nucleon as a bound state of an on-shell ``diquark'' and an off-shell quark.  The full wave function is composed of contributions from both scalar and axial vector diquarks; here we focus on the axial vector contribution only (and we also neglect the isospin factors - for a complete discussion see GRP).  Since the quark is confined, any singularities that might arise from the off-shell quark propagator are cancelled by a zero in the vertex function, and we write the axial vector part of the nucleon wave function as
\ba
\Psi_{\alpha,\lambda_n}(k,P;\varepsilon_p(\lambda))&\equiv& \left<k,\varepsilon_p(\lambda) | q_\alpha(0) | P,\lambda_n\right>\nonumber\\
&=&\varepsilon^{*}_{p\,\mu}(\lambda) \Gamma^\mu_{\alpha\beta}(k,P) u_\beta(P,\lambda_n) \label{eq.9c}\, .\qquad
\ea
The notation differs from Eq.~(\ref{eq5a}).  Here 
$k$ is the four-momentum of an on-shell diquark of mass $m_s$, $q_\alpha(0)$ is the quark field operator with spinor index $\alpha$ (suppressed in the following discussion), $u(P,\lambda_n)$ is a nucleon spinor with four-momentum $P$  and helicity $\lambda_n$, $\Gamma$ is the Dirac structure of the {\it wave function\/} and {\it not\/} the {\it vertex function\/} (it implicitly includes the propagator), and we have included $\varepsilon_p(\lambda)$ in the list of arguments of $\Psi$ for this discussion (but  it is usually suppressed).   Equation~(\ref{eq.9c}) is very similar to (\ref{eq5a}) with the on-shell spinor now describing the bound state (instead of one of the constituents) and the axial vector polarization describing one of the constituents (instead of the bound state).  

 In applications we have chosen the simplest form possible
 \be
 \Gamma^\mu(k,P)=\phi \;\gamma^5\gamma^\mu\, , \label{eq.30}
 \ee
 where $\phi=\phi(P\cdot k)$ is a scalar function of the only variable available (because $k^2=m_s^2$ and $P^2=M^2$), and we suppress the Dirac indices $\alpha,\beta$.  This describes the axial vector diquark contribution to the nucleon wave function, but there is also a scalar diquark contribution and isospin factors, not discussed here.

This matrix element is referred to as the ``wave function'' because it has many of the properties of the solution of a Dirac wave equation.  However, its square is not necessarily a probability density (in common with Klein-Gordon wave functions) so it is not the familiar wave function from quantum mechanics. 

In analogy with last line of Eq.~(\ref{eq5a}), the wave function (\ref{eq.9c}) can be written  as a superposition of all possible spin states of the off-shell quark.  To display this superposition, use the polarization states  (\ref{eq.1}) and the ansatz (\ref{eq.30}) to look at the wave function in the nucleon rest frame, where $P=P_0=\{M,{\bf 0}\}$.   One quickly sees that  the wave function is a four-component spinor that can be written
\ba
\Psi_{\lambda_n}(k,P_0;\varepsilon_0(\lambda))=\sum_{\lambda_q}\psi_{\lambda\lambda_q,\lambda_n}(k,P_0)\left[\begin{array}{c}\chi_{_{\lambda_q}}\cr\cr {\bf 0}\end{array} \right]\, ,\label{eq.8}
\ea
with $\chi_{_{\lambda_q}}=\pm\sfrac12$ and ${\bf 0}$ the two-component  spinors
\ba
\chi_{_{+\frac12}}=\left(\begin{array}{c}1\cr 0\end{array}\right),\qquad \chi_{_{-\frac12}}=\left(\begin{array}{c}0\cr 1\end{array}\right),\qquad{\bf 0}=\left(\begin{array}{c}0\cr 0\end{array}\right)\, , 
\ea
and the spinor index $\alpha$ has been suppressed in favor of the explicit matrix representation of the four components.  Equation (\ref{eq.8}) expresses the quark as a superposition of states with spin ``up'' and ``down''  along the fixed axis.  Furthermore, explicit evaluation of the wave function components
 $\psi_{\lambda\lambda_q,\lambda_n}$ shows that they are zero  {\it unless\/} $\lambda_n=\lambda_q+\lambda$, a consequence of the conservation of angular momentum (because the $\lambda's$ are the spin projections of the particles along the fixed-axis).  The non-zero components of $\psi$  are
\ba
\psi_{\lambda_q\lambda,\lambda_n}(k,P_0) = \phi\times\begin{cases} \;\phantom{-} 1 &\lambda=\phantom{-}0,\; \lambda_n=+\sfrac12, \cr
\;-1 & \lambda=\phantom{-}0,\; \lambda_n=-\sfrac12 \cr
-\sqrt{2} & \lambda=+1,\; \lambda_n=+\sfrac12 
\cr
\phantom{-}\sqrt{2} & \lambda=-1,\; \lambda_n=-\sfrac12 
 \end{cases}
\ea
where $\lambda_q=\lambda_n-\lambda$ in every case.  These are simply a factor of $-\sqrt{3}$ times the familiar vector coupling coefficients $\left<\sfrac12 1\lambda_q \lambda |  \sfrac12 \lambda_n\right>$ for the coupling of  spin 1 and spin $\sfrac12$ states to form a composite state of spin $\sfrac12$.  These results may seem fortuitous, but are only a consequence of how we constructed the wave function in the first place, as discussed in GRP.

The wave function (\ref{eq.9c})  transforms as a spinor.  To show this, first use the properties of the representation $S(\Lambda)$ of the Lorentz transformation $\Lambda$ on the Dirac space, specifically 
%
\ba
&&S(\Lambda)u(P,\lambda_n)=u(\Lambda P,\lambda_{n_W}) \label{eq.9a}\\
&&S(\Lambda) v_\mu \gamma^\mu S^{-1}(\Lambda) = \left(\Lambda v \right)_\mu \gamma^\mu
\ea
%
where here (and in the rest of the paper) we suppress reference to the Dirac indices, $v$ is any four-vector and $\lambda_{n_W}$ is a shorthand notation for the sum
\ba
u(\Lambda P,\lambda_{n_W})\equiv\sum_{\lambda'} u(\Lambda P,\lambda') d_{\lambda'\lambda_n}^{1/2}(R_\Lambda)\, .
\label{Wigr}
\ea
where $d^{1/2}$ are the spin one-half rotation matrices and $R_\Lambda$ is the Wigner rotation induced by the transformation $\Lambda$ (the details of which need not concern us throughout this discussion).  Assuming that $\varepsilon_p^{*\mu}\Gamma_\mu$ is a product of Dirac $\gamma$ matrices contracted with vectors, we obtain 
\ba
S(\Lambda)\Psi_{\lambda_n}(k,P;\varepsilon_p(\lambda))=\Psi_{\lambda_{n_W}}(\Lambda k,\Lambda P;\Lambda \varepsilon_p(\lambda)) \, .\label{eq.10a}
\ea
This is a straightforward generalization of the transformation law (\ref{eq.9a}) for a Dirac spinor.  It is used to connect the rest frame wave function (\ref{eq.8}) to the wave function of the moving nucleon (\ref{eq.9c}).

\subsection{Nuclear current} \label{sec:matrix}

The spectator wave function enters into the evaluation of the matrix element of the nuclear current.   In the simplest approximation (the relativistic impulse approximation which neglects exchange currents) this matrix element is the sum over products of the initial and final state wave functions 
\begin{align}
\left<P_+,\lambda_+|j^\mu(q)|P_-,\lambda_-\right>=&\sum_\lambda\int_k\bar \Psi_{\lambda_+}(k,P_+;\varepsilon_+(\lambda))\nonumber\\
&\times 
j^\mu(q) \Psi_{\lambda_-}(k,P_-;\varepsilon_-(\lambda))\qquad
\label{current}
\end{align}
where the integral is over the momentum $k$ of the noninteracting spectator diquark 
\be
\int_k\equiv \int\frac{d^3k}{(2\pi)^3 2E_k}=\int \frac{d^4k}{(2\pi)^3}\delta_+(m_s^2-k^2)\, ,
\ee
 $P_-$ and $P_+=P_-+q$ are the momenta of the incoming and outgoing nucleons, $\lambda_-$ and $\lambda_+$ are their polarizations,
$E_k=\sqrt{m_s^2+k^2}$ is the diquark on-mass-shell energy
and $\varepsilon_+(\lambda)=\varepsilon_{p_+}(\lambda)$ and $\varepsilon_-(\lambda)=\varepsilon_{p_-}(\lambda)$ are the final and initial state diquark fixed-axis polarization vectors for a diquark state with polarization $\lambda$.  (These vectors must be initially defined in a collinear frame, as dicussed below.)  In the following discussion reference to the nucleon helicities and any Wigner rotations that accompany them will be suppressed; these may always be added to any final formula by using Eq.~(\ref{Wigr}).

Note that this matrix element {\it factorizes\/} in the sense that it is a sum of three terms, each one of which is the {\it product\/} of wave functions.  

The matrix element may be divided into a Dirac part and the polarization sum
\ba
J^\mu=\left<P_+|j^\mu(q)|P_-\right>&=&\int_k \bar u(P_+) \,{\cal A}^\mu_{\nu\nu'}(P_-,q,k) \,u(P_-)\nonumber\\
&&\qquad\times D_{+-}^{\nu\nu'}(P_-,q)\qquad
\label{current2}
\ea
where the Dirac part of the current is
\ba
 &&{\cal A}^\mu_{\nu\nu'}(P_-,q,k)=\bar\Gamma_\nu(k,P_+)j^\mu(q)\Gamma_{\nu'}(k,P_-)\qquad \label{eq.16}
\ea
and the polarization sum is
\ba
D^{\nu\nu'}_{+-}(P_-,q)=\sum_\lambda \varepsilon_+^{\nu}(\lambda)\varepsilon_-^{*\nu'}(\lambda)\, . \label{eq1}
\ea

Evaluation of this spin sum requires careful discussion.  The polarizations are defined with respect to a fixed axis, which we have chosen to lie along the direction of the three momentum of the bound state, so the initial and final polarizations can only be defined consistently in a frame in which the initial and final three momenta of the bound states are collinear.   This is not a restriction, because for {\it any\/} two nucleon four-momenta $P_\pm$ there {\it always\/} exists a Lorentz transformation $\Lambda^{-1}$ that will boost and rotate the two four-momenta so that they lie along the $z$ direction.  Only in this collinear frame are the definitions of the fixed-axis polarizations for the incoming and outgoing diquark  consistently defined with respect to the same $z$ axis, permitting the sum over diquark spins to be evaluated consistently and without error.    Once the sum over the spins has been carried out in this collinear frame, the result can be boosted and rotated back to the original frame using $\Lambda$.    This construction is qualitatively  similar to the procedure used when constructing helicity-states for a two body system.  In that case the construction also starts from the center of mass (a colinear) frame \cite{Jacob:1959at}, and the states are then boosted or rotated to an arbitrary frame.

A simple and elegant way to evaluate the sum (\ref{eq1})  is to exploit the fact that $D_{+-}$ is a sum of direct products of the four-vectors $\varepsilon_+$ and $\varepsilon_-$, and is therefore a covariant tensor.  It can only depend on a sum of bilinear products of  $P^\mu_+$ and $P^\mu_-$ or $g^{\mu \nu}$ (the tensors available).  Using the constraints $P_+\cdot\varepsilon_+=P_-\cdot\varepsilon_-=0$ and $P_\pm^2=M_\pm^2$ (allowing for the possibility that the masses of the incoming and outgoing bound state  are unequal), we see that   $P_{+\mu}D_{+-}^{\mu\nu}=0$ and $D_{+-}^{\mu\nu}P_{-\nu}=0$.  Hence the most general form $D_{+-}$ can have is
\ba
D_{+-}^{\mu\nu}(P_-,q)&=&\sum_\lambda \varepsilon_+^\mu \varepsilon_-^\nu=a_1\left(-g^{\mu\nu}+\frac{P_-^\mu P_+^\nu}{b}\right)\nonumber\\
&&+a_2\left(P_--\frac{bP_+}{M_+^2}\right)^\mu\left(P_+-\frac{bP_-}{M_-^2}\right)^\nu\qquad   
\label{eq.44a}
\ea
where $b=P_+\cdot P_-$.  Using the explicit form (\ref{eq.4}) for the vectors, we see that  $D_{+-}^{xx}=D_{+-}^{yy}=1$ requiring that $a_1=1$.  The coefficient $a_2$ can be found from the trace.  Using the explicit forms (\ref{eq.4}) the trace is
\be
\Big(D_{+-}(P_-,q)\Big)^\mu{}_\mu=-2-\frac{P_+\cdot P_-}{M_+M_-}
\ee
which gives
\be
a_2=-\frac{M_+M_-}{b(M_+M_-+b)}\, .
\ee
For equal masses this reduces to 
\ba
D_{+-}^{\mu\nu}(P_-,q)&=&-g^{\mu\nu}+2\frac{P^\mu P^\nu}{P^2} -\frac{P_+^\mu P_-^\nu}{M^2}\, ,\label{eq3}
\ea
where $P=\sfrac12(P_++P_-)$ and $2P^2=M^2+b$.  This form can also be obtained by explicit construction using the vectors (\ref{eq.4}).  Note that this sum satisfies the covariance condition 
\be
\Lambda^\alpha{}_{\alpha'} \Lambda^\beta{}_{\beta'}
D_{+-}^{\alpha'\beta'}(P_-,q)= D_{+-}^{\alpha\beta}(\Lambda P_-, \Lambda q) \, . \label{eq.22}
\ee

Using (\ref{eq.22}) it is easy to show that the current  (\ref{current}) [or its alternative form (\ref{current2})] is covariant.  This requires proving that it transforms like a four-vector
\be
\Lambda^\mu{}_\nu\, \left<P_+|j^\nu(q)|P_-\right>=\left<\Lambda P_+|j^\mu(\Lambda q)|\Lambda P_-\right> \, .
\ee

Start with the observation that the quark current is a product of scalar functions and gamma matrices, and satisfies the transformation rule
\be
\Lambda^\mu{}_\nu\, j^\nu(q)=S^{-1}(\Lambda) j^\mu(\Lambda q) S(\Lambda) \, .
\ee
Inserting this into the current and using Eq.~(\ref{eq.10a}) gives immediately 
\begin{align}
\Lambda^\mu{}_\nu &\left<P_+|j^\nu(q)|P_-\right>
\nonumber\\
&=\int_k \bar u(\Lambda P_+)\,{\cal A}^\mu_{\alpha\beta}(\Lambda P_-,\Lambda q,\Lambda k)\,u(\Lambda P_-)\nonumber\\
&\qquad\times\left[ \Lambda^\alpha{}_{\alpha'} \Lambda^\beta{}_{\beta'}
D_{+-}^{\alpha'\beta'}(P_-,q)\right]
\, .\qquad
\end{align}
To complete the proof use (\ref{eq.22}) and the fact that the spectator integral over $k$ is covariant
\be
\int_k=\int_{\Lambda k}\, ,
\ee
so that 
\be
\Lambda^\mu{}_\nu \left<P_+|j^\nu(q)|P_-\right>= \left<\Lambda P_+|j^\mu(\Lambda q)|\Lambda P_-\right>\, .
\ee
 As mentioned above, the missing Wigner rotations can be restored by using Eq.~(\ref{Wigr})

In summary: this section has provided a formal proof that both the fixed-axis wave functions and currents are covariant.


\subsection{Critique of the KM discussion of covariance}

 KM discuss what they call ``two interpretations'' of our calculation.  Their ``first interpretation''  is actually a ``misinterpretation'';  it is based on the assumption that we thought the spin sum (\ref{eq1}) could be evaluated in an arbitrary frame using the relation (KM5) [we use the notation KM5 to refer to Eq.~(5) in KM, which is equivalent to our Eq.~(\ref{eq.3}) discussed in the next section.]  We did (and still do) relate the fixed-axis states to helicity states through the rotation (\ref{eq.3}), but we never proposed evaluating the spin sum in an arbitrary frame using this relation.   Eq.~(KM5) {\it can\/} be used to evaluate the spin sum in a collinear frame, as we did in GRP.

We emphasize  that it is essential to start from  a collinear frame in order to properly define the fixed-axis polarization vectors.   It is clear on physical grounds why a collinear frame  is required: only in such a frame are the two fixed-axis polarization vectors $\varepsilon_+$ and $\varepsilon_-$ defined with respect to the {\it same\/} axis, an {\it essential\/} requirement if the fixed-axis sum is to make any sense at all.  (We would have the same problem with a sum of helicity vectors if the angles used to define the initial and final helicity vectors were not defined with respect to the {\it same\/} axis.)  Furthermore, the lab frame is collinear, and hence these frames are very ``natural.''

The ``second interpretation'' discussed by KM is based on the arguments presented in Sec. II above.  These were given to them in the ``advisory'' review.   KM agree that this interpretation is covariant.  So, in spite of the impression created by KM, the issue is not the covariance of the wave functions, but whether or not they have the correct structure (to be discussed below).  There is  {\it only one ``interpretation''\/}, the covariant one we have presented above.

\section{ Comparison with helicity matrix elements}

We now discuss an interesting issue raised by KM and only very briefly addressed by us in our work so far: the comparison of wave functions and currents defined with fixed-axis states to wave functions and currents defined with helicity states.

\subsection{Two definitions of the wave function}

The wave function (\ref{eq.9c}) can be defined for any type of polarization vector.  Using the notation $\xi_k$ to denote any direction-dependent polarization vector with the properties
\ba
k\cdot \xi_k (\lambda)&=& 0\nonumber\\
\xi^*_k(\lambda')\cdot\xi_k(\lambda)&=&-\delta_{\lambda\lambda'} \label{eq.6}\\
\sum_\lambda\xi_k^\alpha(\lambda)\xi^{*\beta}_k(\lambda)&=&-g^{\alpha\beta} + \frac{k^\alpha k^\beta}{m_s^2}\, , \label{sum1}
\ea
the spectator wave function is  (for simplicity, we continue to suppress explicit reference to the nucleon helicities and any Wigner rotations associated with them)
\begin{align}
\Psi_{\lambda_n}(k,P;\xi_k(\lambda))&\equiv \left<k,\lambda | q(0) | P,\lambda_n\right>
\nonumber\\
&=\xi^{*\mu}_k(\lambda){\Gamma'_\mu}(k,P) u(P,\lambda_n) \label{eq.9}
\end{align}
where we continue to suppress all Dirac indices, and the most general form of ${\Gamma'}$ is 
\ba
{\Gamma'_\mu}(k,P)=\gamma^5 \Big[ \phi_1\gamma_\mu+\phi_2P_\mu 
+\phi_3\gamma_\mu\not\! k+\phi_4 \,P_\mu\not\! k \Big]\, . \label{eq.43}
\ea
Since $k^2=m_s^2$ and $P^2=M^2$, each of the scalar functions $\phi_i$ can depend only on the one remaining variable $P\cdot k$, and the Dirac operator can depend only on linear powers in $\not\!k$ and $\not\!P$.  Terms linear in $\not\!P$  can be eliminated using the Dirac equation and  any term dependent on $k^\mu$ will also vanish.  (A similar expansion exists for the fixed-axis wave function, with the $P_\alpha$ terms replaced by $k_\alpha$.)

The remainder of this subsection will be devoted to finding an explicit connection between the fixed-axis wave function  (\ref{eq.9c}) and a {\it specific\/} wave function of the type (\ref{eq.9}).  We will do this in two steps: first we will  construct an explicit relationship between these wave functions in the nucleon rest system, and then we will boost this to a moving frame.  

\subsubsection{Step 1; rest frame connection}

Using $B(k)$ to denote the boost that carries the vector $\{m_s,0,0,0\}$ into $\{E_k,0,0,k\}$ and $R(\theta)$ to denote the rotation through angle $\theta$ about the $y$ axis, helicity-state polarization vectors, denoted by $\xi_{hk}$, are defined by the transformations 
\ba
\xi^\mu_{hk}(\pm)&=&{\cal L}_h^\mu{}_\nu\, \varepsilon^\nu_{0}(\pm)=\mp\sfrac{1}{\sqrt{2}}\left[\begin{array}{c} 0 \cr \cos\theta \cr \pm i \cr -\sin\theta\end{array}\right] \nonumber\\ 
\xi^\mu_{hk}(0)&=&{\cal L}_h^\mu{}_\nu\,\varepsilon^\nu_{0}(0)=\frac{1}{m_s}\left[\begin{array}{c} k \cr E_s\sin\theta \cr 0 \cr E_s\cos\theta\end{array}\right]\, , \label{eq.3}
\ea
where  ${\cal L}_h=R(\theta)B(k)$ is the Lorentz transformation (LT) that converts the $\varepsilon_0$ into helicity vectors.   It is also convenient to introduce ``rotated'' polarization states, denoted by a subscript $r$ and defined by 
\ba
\xi^\mu_{rk}(\lambda)&=&{\cal L}_r^\mu{}_\nu\, \varepsilon^\nu_{0}(\lambda)={\cal L}_h^\mu{}_\nu\,\sum_{\lambda'} \xi^\nu_{0}(\lambda') d_{\lambda'\lambda}(\theta)\nonumber\\
&=&\sum_{\lambda'}\xi_{hk}^\mu(\lambda') d_{\lambda'\lambda}(\theta)\label{eq.3a}
\ea
where ${\cal L}_r=R(\theta)B(k)R^{-1}(\theta)={\cal L}_hR^{-1}(\theta)$ is the LT that converts the $\varepsilon_0$ into {\it rotated\/} vectors $\xi_r$; so named because they are related to helicity vectors by a rotation.

It is possible to write the connection (\ref{eq.3a}) between the vectors $\xi_r$ and $\varepsilon_0$ in a compact manifestly covariant form.   This is because the LT ${\cal L}_r$ is a boost in the direction of ${\bf k}=\{k\sin\theta, 0, k\cos\theta\}$. This boost depends on two four-vectors natural to the problem; the direction of the boost, included in $k^\mu$, and the initial rest frame, in this case the rest frame of the bound state, denoted by $P_0=\{M, 0, 0, 0\}$.  In the rest frame of the bound state, ${\cal L}_r$ can be written
\ba
{\cal L}_r^\mu{}_\nu=g^\mu{}_\nu&-&\frac{m_s}{E_k+m_s}\left(\frac{k^\mu}{m_s} +\frac{P_0^\mu}{M}\right)\left(\frac{k_\nu}{m_s} +\frac{P_{0 \nu}}{M}\right)\nonumber\\
&+&2\left(\frac{k^\mu P_{0\nu}}{m_s M}\right) \label{LT1}
\ea
where the non-covariant looking energy is actually $E_k=k\cdot P_0/M$.  Similarily, the inverse transformation is
\ba
({\cal L}_r^{-1})^\mu{}_\nu=g^\mu{}_\nu&-&\frac{m_s}{E_k+m_s}\left(\frac{k^\mu}{m_s} +\frac{P_0^\mu}{M}\right)\left(\frac{k_\nu}{m_s} +\frac{P_{0 \nu}}{M}\right)\nonumber\\
&+&2\left(\frac{P_0^\mu k_\nu}{m_s M}\right) \label{LT2}
\ea
Using this, and the orthogonality condition (\ref{eq.6}) gives, in the nucleon rest frame,
\ba
\varepsilon^\mu_0(\lambda)=\xi_{rk}^\mu(\lambda)
-\frac{M k^\mu+m_s P_0^\mu}{M(m_s M+P_0 \cdot k)} P_0\cdot\xi_{rk}(\lambda)\, . \qquad\label{eq.12}
\ea

We can use this transformation  to relate the rotated and fixed-axis wave functions (in the rest frame).  Using (\ref{eq.12})  to replace the fixed-axis vectors in (\ref{eq.9c}), we get a wave function of the form (\ref{eq.9}) 
\begin{align}
&\Psi_{\lambda_n}(k,P_0;\varepsilon_0(\lambda))=\varepsilon_0^{*\mu}(\lambda)\Gamma_\mu(k,P_0)\,u(P_0,\lambda_n)\nonumber\\
&\quad=\xi_{rk}^{*\mu}(\lambda)\overset{*}{\Gamma_\mu}(k,P_0)\,u(P_0,\lambda_n)\equiv\overset{*}{\Psi}_{\lambda_n}(k,P_0;\xi_{rk}(\lambda))\qquad \label{eq.26}
\end{align}
where the transformed Dirac operator (with arguments omitted) is
\ba
\overset{*}{\Gamma_\mu}=\Gamma_\mu -\frac{ P_{0\mu}}{m_s M+P_0\cdot k}\left( k\cdot\Gamma+\frac{m_s}{M} P_0\cdot \Gamma\right)\, . \label{eq.20}
\ea 
In the rest frame, the fixed-axis wave function is equal to the wave function with a rotated-state polarization vector, provided the Dirac operator is  transformed according to Eq.~(\ref{eq.20}).

To compare with GRP, [see Eq.\ (\ref{eq.9c}) above], 
we insert the simple operator (\ref{eq.30})  into (\ref{eq.20}) and use the Dirac equation to obtain
\ba
&&\overset{*}{\Gamma_\mu}(k,P_0) u(P_0,\lambda_n)=\nonumber\\
&&\qquad\phi\;\gamma^5 \Big\{ \gamma_\mu -\frac{P_{0\mu}}{M}\frac{\not\!k+m_s}{E_k+m_s}\Big\}u(P_0,\lambda_n)\, .\qquad \label{eq.24}
\ea
Note that $\overset{*}{\Gamma}$ is a special case of the general form (\ref{eq.43}).  

By construction, we know that $\xi^\mu_{rk}\overset{*}{\Gamma_\mu}(k,P_0)u(P_0,\lambda_n)$ does not have any dependence on the angle $\theta$, but it is entertaining and instructive to see how the angular dependence of the individual terms in (\ref{eq.24}) cancel to insure that this is true.  Consider the $\lambda=0$ case. Using the explicit form for $\xi^{*\mu}_{rk}=\xi^{*\mu}_{rk}(0)$
\ba
\xi^\mu_{rk}(0)=\frac{1}{m_s}\left[\begin{array}{c} k  c_\theta\cr(E_s-m_s) c_\theta s_\theta \cr 0 \cr m_s s^2_\theta +E_s c^2_\theta\end{array}\right]\, , \label{eq.3c}
\ea
where $c_\theta=\cos\theta$ and $s_\theta=\sin\theta$,  gives
\begin{align}
&\xi_{rk}^{*\mu}\gamma^5\gamma_\mu u(P_0,\lambda_n)
\nonumber\\
&\qquad=\frac{1}{m_s}\left[\begin{array}{c} (E_s-m_s)c_\theta(\sigma_x s_\theta + \sigma_zc_\theta)
+m_s\sigma_z\cr \cr k\cos\theta \end{array}\right] \chi_{_{\lambda_n}}
\nonumber\\
&\frac{(\xi_{rk}^{*}\cdot P_0)}{M}\gamma^5\frac{\not\!k +m_s}{E_s+m_s} u(P_0,\lambda_n)
= \frac{1}{m_s}\left[\begin{array}{c} {\displaystyle \frac{k\,c_\theta \;{\bf \sigma \cdot k}}{E_s+m_s}}\cr \cr k\cos\theta \end{array}\right] \chi_{_{\lambda_n}}\, ,\quad
\label{eq.38}
\end{align}
where the $\sigma_i$ are the Pauli matrices and $\chi_{_{\lambda_n}}$ is the two component nucleon spinor.  Forming the special combination (\ref{eq.24}) by subtracting the two terms evaluated in (\ref{eq.38})  shows that the lower components cancel, and recalling that ${\bf k}=k\{s_\theta,0, c_\theta\}$ shows that all of the angular dependence in the upper components also cancels, giving the result
\begin{align}
\xi^{*\mu}_{rk}(0)&\overset{*}{{\Gamma}_\mu}(k,P_0) u(P_0,\lambda_n)=\phi\,\left[\begin{array}{c} \sigma_z \cr 0\end{array}\right] \chi_{_{\lambda_n}} \nonumber\\
&=\varepsilon_0^{*\mu}(0)\Gamma_\mu(k,P_0) u(P_0,\lambda_n)\, . \qquad
\end{align}
We have found an explicit way to write a rotated-basis wave function {\it  with no angular dependence\/} in the nucleon rest frame.  This is a pure $S$-state wave function, expressed in terms of polarization vectors satisfying (\ref{eq.6}). 

\subsubsection{Step 2: boosting to a moving frame}

Some care is needed to correctly transform  the correspondence (\ref{eq.26}) from the rest frame where the nucleon has momentum $P_0$ to a moving frame where the momentum is $P$.  This transformation will be represented by the boost Z with the property $P=Z P_0$.  The transformation of the fixed-axis wave function is straightforward
\ba
Z \Psi_{\lambda_n}(k',P_0;\varepsilon_0(\lambda))&=&\Psi_{\lambda_n}(Z k',Z P_0;Z \varepsilon_0(\lambda)) \nonumber\\
&=& \Psi_{\lambda_n}(k,P;\varepsilon_p(\lambda))\label{eq.40}
\ea 
where $Zk'=k$.  (In this discussion $k'$ denotes the momentum in the {\it rest\/} frame, and $k$ the momentum in the boosted frame.)

However, the transformation of helicity or rotated-state wave function involves a Wigner rotation.  In this paper we will develop results for the rotated polarization states defined in Eq.~(\ref{eq.3a}).  These are closely related to helicity amplitudes and, we feel, equivalent for the purposes of this discussion.  We prefer to use them because the formulae are easier to handle, but, with a little more work, all of the results we obtain can also be generalized to helicity states.

Under the boost Z, rotated  polarization vectors  undergo a Wigner rotation
\be
Z^\mu{}_\nu \xi^\nu_{rk'}(\lambda)=\sum_{\lambda'}\xi^\mu_{r\,k}(\lambda') d_{\lambda'\lambda}(\omega_r) \label{eq.9d}
\ee
where the angle $\omega_r$ can be found from the rotation
\ba
R(\omega_r)&=&R(\theta)B^{-1}(k)R^{-1}(\theta)Z R(\theta')B(k')R^{-1}(\theta') \nonumber\\
&=&R(\theta) R(\omega_h) R^{-1}(\theta')\, , \label{eq.10aa}
\ea
where, for comparison, $\omega_h$ is the Wigner rotation angle for helicity states, and $k'=\{E_{k'},k'\sin\theta',0,k'\cos\theta'\}$ are the coordinates of the initial momentum and $k=\{E_{k},k\sin\theta,0,k\cos\theta\}$ are the coordinates of the final, boosted momentum.  As with helicity vectors, the angle $\omega_r$ can be expressed entirely in terms of $P,\,k$, and $\theta$, the final variables after the boost.

Using these relations, the relation (\ref{eq.12}) can
be boosted in the $z$ direction by first replacing $k$ by $k'$ (the rest value of $k$) and then  substituting $k'\to k$, $P_0\to P$ and properly allowing for the Wigner rotation of the polarization state $\xi_{rk}$.  The final result for the boosted vector (\ref{eq.12}) using (\ref{eq.9d}) is
\ba
\varepsilon^\mu_p(\lambda)=\sum_\lambda\Big\{\xi_{rk}^\mu(\lambda')
&-&\frac{M k^\mu+m_s P^\mu}{M(m_s M+P \cdot k)} P\cdot\xi_{rk}(\lambda')\Big\}\nonumber\\
&&\times d_{\lambda'\lambda}(\omega_r)\, . \qquad\label{eq.12b}
\ea
where $\omega_r$ is the Wigner rotation angle defined in Eq.~(\ref{eq.10aa}).  This expression preserves  the properties (\ref{eq.9b}) and (\ref{eq.10b}). 

Using (\ref{eq.12b}), we can now write the boosted wave function (\ref{eq.40}) in terms of the {\it rotated\/} states $\xi_{rk}$
\begin{align}
\Psi_{\lambda_n}\Big(k,P;&\varepsilon_p(\lambda)\Big)=\varepsilon_p^{*\mu}(\lambda)\Gamma_\mu(k,P)\,u(P,\lambda_n)\nonumber\\
&=\sum_{\lambda'}\xi_{rk}^{*\mu}(\lambda')\overset{*}{\Gamma_\mu}(k,P)\,u(p,\lambda_n)\;d_{\lambda'\lambda}(\omega_r)\qquad\qquad
\nonumber\\
&=\sum_{\lambda'}\overset{*}{\Psi}_{\lambda_n}\Big(k,P;\xi_{rk}(\lambda')\Big)\; d_{\lambda'\lambda}(\omega_r)\, ,\label{eq.26a}
\end{align}
where $\overset{*}{\Psi}$ is the wave function (\ref{eq.9}) with the general form $\Gamma'$ replaced by $\overset{*}{\Gamma}$ from Eq.~(\ref{eq.24}).

Note that the two wave functions, $\Psi_{\lambda_n}(k,P;\varepsilon_p(\lambda))$ and $\overset{*}{\Psi}_{\lambda_n}(k,P;\xi_{rk}(\lambda))$ are {\it equal\/} in the nuclear rest frame, but related by a Wigner rotation in the moving frame.

\subsection{Two definitions of the current}

The nuclear current can also be written directly in terms of the rotated polarization states.  In this section we begin by considering the current for the specific wave function $\overset{*}{\Psi}_{\lambda_n}(k,P;\xi_{rk}(\lambda))$ introduced above.  Substituting rotated states $\xi_{rk}$ for the fixed axis states in the current (\ref{current}), substituting the $\overset{*}{\Gamma}$ of (\ref{eq.24}) for the $\Gamma$ in Eq.~(\ref{eq.16}), and replacing the polarization sum $D^{\nu\nu'\beta}_{+-}(p,q)$ by the sum over rotated states, equal to (\ref{sum1}), this current becomes 
\ba
\overset{*}{J_{r}^\mu}&\equiv& \int_k \bar u(P_+,\lambda_+)\, \overset{*}{{\cal A}^\mu_{\nu\nu'}}(P_-,q,k)\, u(P_-,\lambda_-) D_{r}^{\nu\nu'}(k)\qquad\label{currentr}
\ea
where $D_r^{\nu\nu'}$ denotes the polarization sum (\ref{sum1}) and $\overset{\,\,\,*}{{\cal A}}$ is 
\ba
\overset{*}{{\cal A}^{\mu}_{\nu\nu'}}&=& \overset{*}{{\bar \Gamma}_\nu}(k,P_+)\;j^\mu \;\overset{*}{{\Gamma}_\nu'}(k,P_-)
\nonumber\\
&=&\left\{\bar \Gamma_\nu -\left(\frac{P_{+\nu}}{M}\right)\frac{M k\cdot \bar\Gamma+m_s P_+\cdot\bar\Gamma}{m_s M+P_+\cdot k}\right\}\ j^\mu \qquad\nonumber\\
&& \times\left\{ \Gamma_{\nu'} -\left(\frac{P_{-\nu'}}{M}\right)\frac{M k\cdot \Gamma+m_s P_-\cdot\Gamma}{m_s M+P_-\cdot k}\right\} \, . 
\ea
This current is to be compared with (\ref{current2}).  These two currents are very closely related by the following connections: (i) the nuclear wave functions in their rest frames are {\it identical\/} in both cases because of the correspondence (\ref{eq.20}), and (ii) in any frame, both wave functions satisfy the Dirac equation $(\not\!\! P-M)\Psi=0$.   In applications, we chose  the $\Gamma$ defined in Eq.~(\ref{eq.30}).  This gives a wave function with a {\it pure\/} $S$-wave structure in the nuclear rest frame, the only frame where the discussion of nuclear shape makes any sense.

The differences between these two currents [(\ref{current2}) and (\ref{currentr})] is subtle, and helps to clarify the issues raised by KM.  
To explain these differences, we write the fixed-axis current (\ref{current2}) in an alternate form using the correspondences (\ref{eq.26a}).  Substituting (\ref{eq.26a}) into the fixed-axis current (\ref{current2}) gives
\begin{align}
\tilde J^\mu\equiv& \int_k \bar u(P_+,\lambda_+)\, {\overset{*}{ {\cal A}^\mu_{\nu\nu'}}}(P_-,q,k)\, u(P_-,\lambda_-) \nonumber\\
&\qquad\times
\overset{*}{D_{r}^{\nu\nu'}}(P_-,q,k)\,  ,\quad \label{eq.51}
\end{align}
where  the polarization sum is
\ba
\overset{*}{D^{\nu\nu'}_r}(P_-,q,k)&\equiv&\sum_{\lambda\lambda_f\lambda_i}\!\!\!\! \xi_{rk}^\nu(\lambda_f)\xi^{*\nu'}_{r\, k}(\lambda_i)\nonumber\\
&&\qquad\times d_{\lambda_f\lambda}(\omega_{+}) d_{\lambda_i\lambda}(\omega_{-})\, , \label{eq.47a}
\ea
and $\omega_\pm$ are the Wigner rotations for the transformations from the rest frame to the initial and final nucleon momenta, $P_\pm$.   We emphasize that the two forms of the current, (\ref{current2}) and (\ref{eq.51}), are {\it identical\/}
\be
\tilde J^\mu=J^\mu
\ee

Comparing  (\ref{eq.51}) with the rotated current,  (\ref{currentr}), we see that the difference is the replacement of the sum $D_r(k)$ by the sum $\overset{*}{D_r}(P_-,q,k)$.  From the rotated state point of view, the spin sum $\overset{*}{D}$ describes a case in which  {\it the spin of the spectator diquark is transformed\/} from $\lambda_i$ to $\lambda_f$ as it propagates from the incoming nucleon to the final nucleon, and is a departure from  the impulse approximation, where the spin projection should remain unchanged.

Alternatively, if we want to preserve the impulse approximation exactly, we can use the current (\ref{currentr}).  This will have the most important  features of the fixed-axis current (\ref{current2}); in particular, it describes states which are pure $S$-waves in the rest frame.

We have found that the fixed axis current has many advantages.  It not only provides a good phenomenology for describing the nucleon form factors, but it also allows us to construct orthogonal $N$ and $\Delta$ wave functions, giving a description of  the $\gamma^*+N\to\Delta$ transition which is gauge invariant in a natural way.  Furthermore, starting from simple pure $S$-state wave functions we can add, by explicit construction, angular momentum components $L>0$ to the wave function and study their influence.  We find that these $L\ne0$ components are essential for a full description of the three form factors that enter the $\gamma^*+N\to\Delta$ transition, providing relativistic confirmation of results very familiar from nonrelativistic theory.   These results will be described in Ref.~\cite{GRP2}.   It remains to be seen whether or not the current (\ref{currentr}) will be as effective; this is planned for future study.

\subsection{Critique of the KM discussion of these issues}

In Sec.~IV of their paper, KM obtain Eq.~(KM26), which is identical to our final result (\ref{current2}) for the fixed-axis current.  They  say that this equation should be compared to the result that would be obtained using helicity-state polarizations, and we agree and we have done this.  We found a form for the wave function in a rotated basis that {\it exactly\/} reproduces our fixed-axis wave function.  Our results are (\ref{eq.26}) (rest frame) and (\ref{eq.26a}) (moving frame).  In this respect, the two approaches are equivalent. 

This comparison can be extended to the currents, where we find a difference.  The fixed-axis current (\ref{current2}) can be {\it exactly\/} transformed into the rotated basis, giving Eq.~(\ref{eq.51}).  This differs from the result we would obtain if we started directly from the rotated basis, which would give Eq.~(\ref{currentr}).  The difference between (\ref{eq.51}) and (\ref{currentr}) is summarized by the replacement of the  polarization sum $D_r$ from Eq.~(\ref{sum1}) with the sum $\overset{*}{D}_r$, from Eq.~(\ref{eq.47a}).  In the sum $\overset{*}{D}_r$ the spin projection of the diquark undergoes a Wigner rotation as it propagates from the initial to the final state, and hence, from the viewpoint of the rotated basis, is not propagating as a free particle, and the approximation is not an impulse approximation.  So we agree with KM that the fixed-axis cannot be written exactly in the form (\ref{currentr}).  The fixed axis current includes some additional interactions not of the impulse form.

However, it is not clear how physical important this difference really is. Both currents [(\ref{eq.51}) and (\ref{currentr})] correspond to nuclear wave functions with a pure $S$-state structure  (provided, of course, that the specific form (\ref{eq.24}) is used for the Dirac structure of the rotated wave function), a feat that KM seem to believe is impossible.    

In addressing this comparison,  KM raise several issues that they seem believe are serious shortcomings of the fixed-axis approach.  In our language, these are:

\begin{itemize}

{\item The integrand in (\ref{current2})  is not factorizable into products of independent wave functions because of factors like $2P^2=M^2+P_+\cdot P_-$ in the denominator of the spin sum $D_{+-}$; } 

{\item Evaluation of the current in a simple case [as illustrated in (KM30)] will not give terms in the numerator that will cancel the term $2P^2=M^2+P_+\cdot P_-$ in the denominator of the spin sum $D_{+-}$; }

{\item The transformation $\Lambda$ that carries the current from an arbitrary frame to a colinear frame depends on both $P_+$ and $P_-$ and hence it is misleading to say that the polarization vectors $\varepsilon_+$ ($\varepsilon_-$) [used in (\ref{eq.44a})]  depend only on $P_+$ ($P_-$) only;  }

\end{itemize}

All of these objections can be dealt with easily.  
First, the integrands of both (\ref{current2}) and ({\ref{currentr}) 
clearly display the current as a sum of the product of three terms 
(one for each polarization projection $\lambda$).}  
Furthermore, the presence of a term $P^2$ in the denominator of the spin sum is natural and expected.  While such terms appear immediately in the fixed-axis sum, they also arise eventually from any direction-dependent polarization sum (\ref{sum1}).   For example,  a ``factorizable'' term like
\ba
\gamma_\alpha\gamma^\mu \gamma_\beta\; (k^{\alpha} k^{\beta} )= 2k^\mu\not\!k-\gamma^\mu m_s^2\label{kkterm}
\ea
that is part of any calculation of the current (\ref{currentr}), must be averaged over the directions of ${\bf k}$ (before integrating over $k$).  The angular  integral is most easily  evaluated in the Breit frame (which can always be chosen because the current is covariant).   Assuming the product of the wave functions will be even in $\cos\theta$ (true 
for any form factor in the Breit frame) we may use the nice identity to evaluate the integral over the azimuthal angle $\phi$ 
\be
\int_0^{2\pi} \frac{d\phi}{2\pi}\, k^\mu k^\nu = ag^{\mu\nu}+b\frac{P^\mu P^\nu}{P^2}+c\frac{q^\mu q^\nu}{Q^2}\, .
\ee
The term in (\ref{kkterm}) bilinear in $k$ therefore generates contributions of the form  
\be
\int_0^{2\pi} \frac{d\phi}{2\pi}\,  k^\mu \not\!k=a\gamma^\mu +b \frac{P^\mu \not\!P}{P^2} +c \frac{q^\mu \not\!q}{Q^2}\, ,
\ee
which includes a term with $P^2$ in the denominator.  These factors did indeed appear in Ref.~\cite{Gross06} [see Eqs.~(19 and (20)] where the diquark polarization was defined using the helicity basis.

The third objection overlooks the fact that the fixed-axis polarization vectors are defined {\it first\/} in a collinear frame and {\it then\/} boosted to an arbitrary frame.  This means that even if we start from an arbitrary frame, we must first boost to a collinear frame before we can define the polarization vectors.  If the initial momenta are $P_+'$ and $P_-'$, and the final state collinear vector is $P_+=\Lambda(P_+',P_-')P_+'$, the fixed-axis polarization vector $\varepsilon_+$ will depend on $P_+'$ and $P_-'$ {\it only through\/} the vector $P_+$, which is no different than it would have been if we started from the collinear frame.

We wish to take issue with a number of other statements in the KM paper.  They say (for example) that the boost ${\cal L}_k^{-1}$ in the direction $-{\bf k}$ is not a covariant tensor, and in Eq.~(KM14) they write it in a non-covariant form with explicit factors like $\delta^{\mu0}$.  But the properties of the Lorentz group tell us how to  transform it from from one frame to another, and it will be covariant (invariant in form) if we can find a way to write it in terms of the physical vectors available to us.

This is precisely what we did in this paper, and the technique gave the very nice correspondence (\ref{eq.12}).   In this problem we have two vectors available, $k$ and the momentum vector for the bound state at rest, which can be written $P^\mu=\delta^{\mu 0}ÊM \equiv P_0^\mu$ (we all know that this does not mean that $P^\mu$ is not a four-vector).   If there were only one vector available, the boost ${\cal L}_k^{-1}$ could not be written in a covariant form because there would be no way to characterize the factors $\delta^{\mu0}$.   But the second vector, $P_0$, enabled us to expressÊ the boost in the rest frame of the bound state by replacing $\delta^{\mu0}$ with $P^\mu_0$.  Once this had been done, the expression could be transformed to an arbitrary frame by further LT's.  

We do not think rotations about the $y$ axis can be treated in this way because there is no physical vector defining the $y$ direction (unless we were to introduce a polarized nucleon with spin in the $y$ direction, but this would pose other probelms, we think).  This was the reason we introduced rotated polarization vectors -- to avoid explicit references to the rotations.  
However, avoiding rotations is not really necessary; their transformation properties can be handled in the same way we treat Wigner rotations, but the formulae we obtain are less elegant.

\section{Conclusions}

To describe the polarization of a diquark bound to a nucleon, we proposed, in GRP, to use new, so called ``fixed-axis'' polarization states instead of the usual direction-dependent polarization states (either helicity or the ``rotated'' states defined here).  The latter depend on the diquark momentum $k$, and therefore on its direction, satisfying $\xi_{k}\cdot k=0$.  In fact, in the rest frame of the nucleon, both choices can be made exactly equivalent through an appropriate redefinition of the vertex function $\Gamma$.  
Then, in the case of a pure $S$-state wave function 
(easy to construct using the fixed-axis states) 
the transformation from diquark fixed-axis polarization states to 
direction-dependent states  gives a vertex function with  just the right angular dependence on the diquark momentum to cancel the dependence introduced by the direction-dependent states. 
Conversely, a spherically symmetric vertex function, if taken together with the direction dependent diquark states would result in a wave function without spherical symmetry.  Since we want to investigate the consequences of using wave functions that are spherically symmetric, it is natural to write these wave functions in terms of fixed-axis diquark polarizations. 

KM have strongly criticized this approach, and this paper not only develops new ideas not yet published, but also answers most of the KM objections.  In particular, we have proved that use of fixed-axis polarization states is covariant (provided they are defined correctly), and have shown explicitly how vertex functions and currents using fixed-axis states can be transformed into martix elements involving direction-dependent states.  We have shown, as a consequence, that it is possible, using any type of polarization state, to construct a covariant spherically symmetric nuclear wave function.  

\vspace{0.3in}

This work was partially support by Jefferson Science Associates, LLC under U.S. DOE Contract No. DE-AC05-06OR23177.  G.\ R.\ was supported by the portuguese Funda\c{c}\~ao para a Ci\^encia e Tecnologia (FCT)
under the grant SFRH/BPD/26886/2006.

\end{document}